\documentclass[12pt]{article}
\newcommand{\beao}{\begin{eqnarray*}}
\newcommand{\eeao}{\end{eqnarray*}}
\newcommand{\be}{\begin{equation}}
\newcommand{\ee}{\end{equation}}
\newcommand{\bea}{\begin{eqnarray}}
\newcommand{\eea}{\end{eqnarray}}
\newcommand{\beq}{\begin{eqnarray}}
\newcommand{\eeq}{\end{eqnarray}}
\newcommand{\nn}{\nonumber}
\newcommand{\pa}{\partial}
\newcommand{\ep}{\epsilon}
\newcommand{\al}{\alpha}
\newcommand{\la}{\lambda}
\newcommand{\Ref}[1]{(\ref{#1})}
\newcommand{\st}{\sinh(t)}
\newcommand{\cs}{\cosh(s)}\newcommand{\ct}{\cosh(t)}

\usepackage{epsfig}%\input{psfig}

\def\be{\begin{equation}}
\def\ee{\end{equation}}
\def\bea{\begin{eqnarray}}
\def\eea{\end{eqnarray}}
\def\l{\label}
\def\b{\beta}

\def\o{\over}

\def\ep{\epsilon}

\def\G{\Gamma}

\def\qq{\qquad}

\def\cqq{\;, \qquad}

\begin{document}
\title{Non-transversality of the gluon polarization tensor in a chromomagnetic background}
\author{
{\sc M. Bordag}\thanks{e-mail: Michael.Bordag@itp.uni-leipzig.de} \\
\small  University of Leipzig, Institute for Theoretical Physics\\
\small  Vor dem Hospitaltore 1, 04103 Leipzig, Germany\\ [8pt]
{\sc J. Grebenyuk}\thanks{e-mail: iuli@mail.ru} \\
\small  St.Petersburg State University, \\ \small Universitetskaja nab. 7/9, 199034 St.Petersburg, Russia \\
\small and\\
{\sc V. Skalozub}\thanks{e-mail: Skalozub@ff.dsu.dp.ua}\\
\small  Dnepropetrovsk National University, 49050 Dnepropetrovsk, Ukraine}
 \maketitle
%\tableofcontents
 \begin{abstract}
 We investigate the question about the transversality of the gluon polarization tensor in a homogeneous chromomagnetic background field. 
We re-derive the non transversality known from a pure one loop calculation    using the Slavnov-Taylor identities. In addition we generalize the procedure to arbitrary gauge fixing parameter $\xi$ and calculate the $\xi$-dependent part of the  polarization tensor.
 \end{abstract}
 \section{Introduction}
Recently the question about the structure of the quark-gluon plasma at high temperature attracted new attention. In both, perturbative \cite{Skalozub:2002da,Skalozub:2004ab,Starinets:1994vi,Skalozub:1999bf,Demchik:2002ks} and lattice \cite{Agasian:2003yw} calculations it was found that a state with a spontaneously generated magnetic background field is privileged. This is interesting especially in view of the common opinion about a gas of free gluons (and quarks) in the deconfinement phase. In general, in non-abelian gauge theory it was observed long ago that a magnetized vacuum has lower energy as the perturbative one  and that this state is, however, not stable due to the tachyonic mode. The recent achievement is to make resummations in the perturbative approach at high temperature where it was found that the gluons acquire a magnetic mass which removes the instability (at least in the approximations considered). This is a motivation for the investigation of the gluon polarization tensor in a magnetic background field.

The  polarization tensor in a magnetic field with and without temperature is the basic object needed in all kinds of resummations in the perturbative approach. Its investigation has a long history, especially in QED. In this paper we focus on the question about its transversality, i.e., whether the relation $p_\mu\Pi_{\mu\nu}(p)=0$ holds. In abelian theories in a background field with and without temperature its transversality is out of question. In QCD its transversality was established at finite temperature but without magnetic background. It is a common opinion that it is transversal in a magnetic background too. However, this turns out not to be the case.  In \cite{Bordag:2005br} its non-transversality in a magnetic background field was shown in a pure one-loop calculation at zero temperature. Only the weaker condition $p_\mu\Pi_{\mu\nu}(p)p_\nu=0$ holds.

The non-transversality has obviously consequences for the gluon spectrum in a magnetic field and for the high temperature plasma. This motivates a detailed investigations of the non-transversal part. In \cite{Bordag:2005br} it was found  on the pure one loop level from the calculation of the corresponding graphs. In the present paper  we derive it from the Slavnov-Taylor identities. At once we investigate the dependence of the polarization tensor on the gauge fixing parameter.

In this paper w use the same notations as in \cite{Bordag:2005br}, for instance we put the constants $\hbar$, $c$ and $g$ equal to unity and we write all formulas in Euclidean notation.

%%%%%%%%%%%%%%%%%%%%%%%%%%%%%%%%%%%%%%%%%%%%%%%%%%%%%%%%%%%%%%%%%%%%%%%%%%%%%%
%%%%%%%%%%%%%%%%%%%%%%%%%%%%%%%%%%%%%%%%%%%%%%%%%%%%%%%%%%%%%%%%%%%%%%%%%%%%%%
%%%%%%%%%%%%%%%%%%%%%%%%%%%%%%%%%%%%%%%%%%%%%%%%%%%%%%%%%%%%%%%%%%%%%%%%%%%%%%
\section{Slavnov-Taylor identity}
In this section we adopt the Slavnov-Taylor identity to show the non-trans\-ver\-sality of the gluon polarization tensor. We consider SU(2) gluodynamics in a homogeneous chromomagnetic background field. The initial gauge potential, $A_\mu^a$, where $a=1,2,3$ is the color index, is splitted into the background, $B_\mu^a$, and the quantum fluctuations, $Q_\mu^a$, according to $A_\mu^a=B_\mu^a+Q_\mu^a$. The background chromomagnetic field points in the third direction in both color  and coordinate space and its potential can be chosen as $B_\mu^a=\delta^{a3}\delta_{\mu 1}x^2$. We work in Euclidean space and the Lagrange density takes the conventional form
\begin{equation} \label{lagrangian}
L = - \frac{1}{4} F^a_{\mu \nu} F^a_{\mu \nu} + \frac{1}{2 \xi} (D_{\mu} Q_{\mu})^2 - \bar{\eta}^a M^{a
b }\eta ^b ,
\end{equation}
where $\eta^a$ is the ghost field.
Here all derivatives are covariant ones with respect to the background field,
\be\label{cder}D^{a b}_{\mu} = \delta^{ab}\frac{\partial}{\partial x^{\mu}} +  g \epsilon^{a s b}B^s _{\mu}
\ee
and they are related to its field strength $B^{s}_{\mu \nu}$ in the usual way, $[D_{\mu}, D_{\nu}]^{a b} = \epsilon^{asb}B^{s}_{\mu \nu}$.
The substitution of the derivatives holds for instance for the gauge fixing term ($\xi$ is the corresponding gauge fixing parameter)  and for the kernel of the ghost term,
\be \label{defM}M^{a b} =
D^{a c }_{\mu}( D^{c b}_{\mu} + \epsilon^{c s b }Q^s _{\mu} ).
\ee
We remark that in this way the magnetic background field is introduced just in the same way as in the well known background field method \cite{Abbott:1981ke}. For instance, from that method it is known that the background can be added by formally substituting all occupancies of the  ordinary derivative by the covariant one \Ref{cder}. In this way the basic notations are defined. For all further definitions and notations we refer to the paper \cite{Bordag:2005br}.

We are going to derive the non-transversality of the polarization tensor from the Slavnov-Taylor identities. Frequently, these are used in the framework of the BRST formalism which is the most efficient way to show the renormalizability of a non-abelian gauge theory. However, in our case we do not need the full formalism and can restrict ourselves to an easier formulation as, for example given in the book \cite{Faddeev1980as}. We start with formula (7.14) in chapter 4 there. After conversion to Euclidean space and written in our notations it reads
\begin{equation}\label{WI1}
\left\langle \frac{1}{\xi} D_{\nu}^{bb'}(y) Q^{b'}_\nu(y)+\int dx \
J^a_\mu(x) (D_\mu^{ac}(x) +  \epsilon^{a s c} Q_{\mu}^s(x) )
(M^{-1})^{cb}(x,y) \right\rangle = 0,
\end{equation}
where we denoted the dependencies on the coordinates $x$ and $y$ explicitly.
We remark that \Ref{WI1} in \cite{Faddeev1980as} was derived without background field. However, it is obvious that this formula holds also with background field if  the derivatives are substituted by the covariant ones.

In \Ref{WI1} the average denotes the corresponding functional integration over $Q_\mu^a$. The specific identity we are interested in for the polarization tensor can be obtained by taking the functional derivative with respect to the source $J_\mu^a(x)$ and putting $J=0$ afterwards. In this way we obtain
\begin{equation}\label{WI2}
\left\langle \frac{1}{\xi} Q_\mu^a(x)
D_\nu^{bb'}(y)Q^{b'}_\nu(y)+\left(D_\mu^{ac}(x)+\epsilon^{asc}Q_\mu^s(x)\right)
{(M^{-1})}^{cb}(x,y)\right\rangle =0.
\end{equation}
Next we need to consider the functional average. For the two point functions it holds
\be\label{fa}
 \left\langle Q_\mu^a(x)Q_\nu^b(y)\right\rangle= G_{\mu\nu}^{ab}(x,y)
-G_{\mu\lambda}^{as}(x,z)\Pi_{\lambda\lambda'}^{st}(z,z')G_{\lambda'\nu}^{tb}(z',y)+\dots .
\ee
Here and in the following integration over $z$ and $z'$ is assumed. In \Ref{fa} $G_{\mu\nu}^{ab}(x,y)$ is the free Greens function of the gluon obeying
\begin{equation}\label{freeprop1}
\left(- \delta_{\mu\lambda}(D^2)^{ac}(x)+
\left(1-\frac{1}{\xi}\right)D_{\mu}^{aa'}(x)D_{\lambda}^{a'c}(x)-
2 \epsilon^{asc}B_{\mu\lambda}^{s}\right) \
G_{\lambda\nu}^{cb}(x,y)=\delta(x-y)\delta^{ab}\delta_{\mu\nu}
\end{equation}
and $\Pi_{\lambda\lambda'}^{st}(z,z')$ is the polarization tensor, i.e., the sum of all 1PI diagrams with two external legs. The second term in \Ref{fa} can be expanded into powers of $Q_\mu^a$ according to
\begin{eqnarray} \label{M-1} (M^{-1})^{cb}(x,y)&=&
G^{cb}(x,y)
+  G^{c c''}(x,z)D_\lambda^{c''c'}(z)\epsilon^{c'sb'}Q_\lambda^s(z)G^{b' b}(z,y) \\ [8pt]
&&\hspace{-2cm}+
G^{c c''}(x,z)D_\lambda^{c''c'}(z)\epsilon^{c'sd}Q_\lambda^s(z)G^{d d''}(z,z')D_{\lambda'}^{d''d'}(z')
\epsilon^{d'tb'}Q_{\lambda'}^t(z')G^{b' b}(z',y) +\dots . \nonumber
\end{eqnarray}
where $G^{ab}(x,y)$ is the free Greens function of the ghosts obeying
\be\label{freeprop2}(D^2_x)^{ac} \ G^{cb}(x,y)=-\delta^{ab}\delta(x-y).
\ee
Into \Ref{M-1}  arbitrary powers of $Q$ enter. Now we take into account that we are interested in an identity for the one-loop polarization tensor, i.e., in the second order in the coupling $g$ in Eq. \Ref{WI2}. We remark that the counting of the powers of $g$ includes only that which go with the field $Q$, not that which we included into the background field $B$. For this reason we need to collect the terms which are second order in $Q$ from Eq.\Ref{WI2}. In the second term in \Ref{WI2} this results in
\bea\label{WI3}
&&\left(D_\mu^{ac}(x)+\epsilon^{asc}Q_\mu^s(x)\right)
{(M^{-1})}^{cb}(x,y)
\\&&=
\dots+
D^{ac}_\mu(x)   G^{cd}(x,z) D^{de}_\nu(z) \ \epsilon^{esf} Q^s_\nu(z)  G^{fh}(z,z') D^{hi}_\lambda(z') \ \epsilon^{itj}Q^t_\lambda(z') G^{jb}(z',y)    \nn \\
&&+\ \epsilon^{asc} Q^s_\mu(x)  G^{cd}(x,z) D^{de}_\lambda(z) \ \epsilon^{etf}Q^t_\lambda(z) G^{fb}(z,y) +\dots   \nn  \\
&& = \left(\delta^{ae}\delta_{\mu\nu}\delta(x-z)+D^{ac}_\mu(x)   G^{cd}(x,z) D^{de}_\nu(z)\right)
\Sigma^{ef}_\nu[Q](z,z')G^{fb}(z',y)
\eea
with
\be\label{SiQ}\Sigma^{ef}_\nu[Q](x,z)= \ \epsilon^{esf} Q^s_\nu(z)  G^{fh}(z,z') D^{hi}_\lambda(z') \ \epsilon^{itj}Q^t_\lambda(z').
\ee
Now we take the functional average. In the first term in Eq.\Ref{WI2} this delivers to order $g^2$ the second term from the r.h.s. of \Ref{fa} where $\Pi^{st}_{\lambda\lambda'}(z,z')$ must be taken in one-loop order and in the second term in \Ref{WI2}, rewritten in the form \Ref{WI3}, the given order comes from the first term in the average \Ref{fa},
\be\label{avSi}\left\langle \Sigma^{ef}_\nu[Q](x,z)\right\rangle
\equiv \Sigma^{ef}_\nu(x,z)= \ \epsilon^{esf}   G^{fh}(z,z') D^{hi}_\lambda(z') \ \epsilon^{itj}G^{ts}_{\lambda\nu}(z',z).
\ee
Taking this together we get
\bea\label{WI4}
&&\frac{1}{\xi}G^{as}_{\mu\lambda}(x,z)\Pi^{st}_{\lambda\lambda'}(z,z')
G^{tc}_{\lambda'\nu}(z',y)D^{cb}_\nu(y)
\nn \\ &&=-\left(\delta^{ae}\delta_{\mu\nu}\delta(x-z)+D^{ac}_\mu(x)   G^{cd}(x,z) D^{de}_\nu(z)\right)
\Sigma^{ef}_\nu(z,z')G^{fb}(z',y),
\eea
where the derivative $D^{bb'}_\nu(y)$ acting in \Ref{WI2} on $Q^{b'}_\nu(y)$ had been integrated by parts so that it acts now to the right (one may assume test functions supplemented). Next we use the obvious properties of the gluon Greens function
\be\label{prop1}G^{ac}_{\mu\lambda}(x,y)D^{cb}_\lambda(y)=\xi D^{ab}_\mu(x)G(x,y) \ee
which is the covariant generalization of the simple momentum space relation
$$
\left(\frac{\delta_{\mu\lambda}}{p^2}-(1-\xi)\frac{p_\mu p_\lambda}{p^4}\right)p_\lambda=\xi\frac{p_\mu}{p^2}.
$$
Of course, in \Ref{prop1} the order of factors matters. Using \Ref{prop1} the explicit dependence on the gauge fixing parameter $\xi$ disappears in the l.h.s. of Eq.\Ref{WI4}. Next we use \Ref{freeprop1} and multiply Eq.\Ref{WI4} from the left with the operator in \Ref{freeprop1}. On the l.h.s. all factors in front of $\Pi$ disappear, in the r.h.s. we get
\bea\label{auxrel}
&& \left( -(D^2_x)^{sa}\delta_{\lambda\mu}-2\epsilon^{sta}B^t_{\lambda\mu}(x)
+\left(1-\frac{1}{\xi}\right)D^{su}_\lambda(x)D^{ua}_\mu(z)\right)
\nn \\ && \times \left(\delta^{ae}\delta_{\mu\nu}\delta(x-z)+D^{ac}_\mu(x)   G^{cd}(x,z) D^{de}_\nu(z)\right)
\nn \\ &&
=-(D^2_x)^{se}\delta_{\lambda\nu}-2\epsilon^{ste}B^t_{\lambda\nu}(x)+D^{sc}_\lambda (x) D^{ce}_\nu(x)
\nn \\ &&
\equiv K^{se}_{\lambda\nu}(x)\eea
where the commutator relation for the covariant derivative was used. Finally we remove the Greens function $G(y,z)$ on the right of both sides by making use of  $G(y,z)\pa_z^2=\delta(y-z)$. In this way we arrive at
\be\label{Wicoord}
\Pi^{st}_{\lambda\lambda'}(z,z')D^{tu}_{\lambda'}(z')=K^{st}_{\lambda\lambda'}(z) \Sigma^{tu}_{\lambda'}(z,z')
\ee
with $\Sigma^{tu}_{\lambda'}(z,z')$ given by \Ref{avSi}.
The r.h.s. of this equation describes the non transversality of the polarization tensor.
It is clear that by virtue of hermiticity a similar relation holds with the covariant derivative acting on the left side of the polarization tensor. Further we remark that from $D^{ab}_\mu(x)K^{bc}_{\nu\lambda}(x)=0$ the 'weak transversality'
\be
\label{weak} D^{as}_\lambda(z)\Pi^{st}_{\lambda\lambda'}(z,z')D^{tb}_{\lambda'}(z')=0
\ee
follows.

The question whether the polarization tensor is transversal or not can now be answered by the investigation of $\Sigma$. To do this for a homogeneous chromomagnetic background field it is meaningful to transform to the so called 'charged basis', where all quantities $f^a$ carrying a color index are transformed according to
\be\label{tracha} f^\pm=\frac{1}{\sqrt{2}}\left(f^1\pm i f^2\right), \qquad f=f^3 .\ee
A summation over doubly appearing color indices reads than
$$ f^a f^a=f^-f^++f^+f^-+ff .$$
The interpretation is that $f$ is color neutral and the $f^\pm$ carry positive resp. negative color charge. With respect to the algebraic structure this is in complete analogy with the electric charge. As a result, a color neutral field does not interact with the background whereas the color charged fields do (rotating on their Landau levels in opposite directions). The epsilon tensor remains antisymmetric and is normalized according to $\epsilon^{3-+}=i$.

Applying this transformation  we obtain an identity for the polarization tensor of the neutral gluons,
\be\label{WIn} \Pi_{\lambda\lambda'}(z,z')\frac{\partial}{\partial {z'}^{\lambda'}}=
K^{33}_{\lambda\lambda'}(z,z')\Sigma^{33}_{\lambda'}(z,z')
\ee
and one for the polarization tensor of the charged gluons,
\be\label{Wich} \Pi^{-+}_{\lambda\lambda'}(z,z')D^{-+}_{\lambda'}(z')=
K^{-+}_{\lambda\lambda'}(z,z')\Sigma^{-+}_{\lambda'}(z,z').
\ee
We note that the 'neutral' covariant derivative is just an ordinary one, $D^{33}_\mu(x)=\frac{\partial}{\partial x^\mu}$, whereas the 'charged' derivative reads
$$ D^{-+}_{\mu}(x)\equiv D_\mu(x)=\frac{\partial}{\partial x^\mu}+iB_\mu(x). $$
For the neutral component of $\Sigma$ we get
\be\label{Sin} \Sigma^{33}_{\lambda'}(z,z')=-2 \Re \ G^{-+}(z,z')D_{\lambda'}(z')G^{-+}_{\lambda'\lambda}(z',z),
\ee
where $\Re$ denotes the real part.
Initially there are  2  contributions in $\Sigma^{33}$. They are complex conjugated one to the other.
For the charged component we get
\be\label{Sich} \Sigma^{-+}_{\lambda'}(z,z')=-G^{-+}(z,z')D_{\lambda'}(z')G_{\lambda'\lambda}(z',z)-
G(z,z') \frac{\partial}{\partial {z'}^{\lambda'}}G^{-+}_{\lambda\lambda'}(z,z').
\ee
We note the symmetry property $G^{-+}_{\lambda\lambda'}(z,z')=G^{+-}_{\lambda'\lambda}(z',z)$.
In \Ref{Sin} and \Ref{Sich} $G$, $G_{\lambda\lambda'}$ and $G^{-+}_{\lambda\lambda'}$
are the Greens functions of a neutral scalar, of a charged scalar and of a charged vector accordingly.

Finally we rewrite these relations in momentum space representation. We use the notations
$$ \Pi_{\lambda\lambda'}(x,y)=\int\frac{dp}{(2\pi)^4} \ e^{-ip(x-y)} \ \Pi_{\lambda\lambda'}(p) $$
and
$$ \Sigma_{\lambda}(x,y)=-i \int\frac{dp}{(2\pi)^4} \ e^{-ip(x-y)} \ \Sigma_{\lambda}(p) . $$
As a convention we preserve the notation $p$ for the momentum of a 'charged function' and $k$ for a neutral one. In this way   we write
\be\label{Gp} G^{-+}(x,y)=\int\frac{dp}{(2\pi)^4} \ e^{-ip(x-y)} G(p), \ee
\be\label{Gk} G^{}(x,y)=\int\frac{dk}{(2\pi)^4} \ e^{-ik(x-y)} G(k) \ee
and similar for $\Pi$ and $\Sigma$.

Of course, the momentum representation for a charged particle in a magnetic field needs a comment. We assume the application of Schwinger's operator method \cite{Schwinger:1973kp}, where $p_\mu$ is the Fourier transform of the covariant derivative \Ref{cder} and it is an operator,
$$ iD^{-+}_\mu(x) \ \to \ \hat{p}_\mu=p_\mu-B_\mu(i\frac{\partial}{\partial p}) .$$
In the following we drop the hat and use the fact that all formal relations in the transformation into momentum representation stay in place provided the non commutativity of the components of $p_\mu$, $[p_\mu , p_\nu]=iB_{\mu\nu}$ is taken into account. Note that in these notations the momentum $k_\mu$ is commuting. Using these rules we obtain for the polarization tensor of the neutral gluons
\be\label{Pik} \Pi_{\lambda\lambda'}(k)k_{\lambda'}=K_{\lambda\lambda'}(k)\Sigma_{\lambda'}(k)
\ee
with \be\label{Kk}  K_{\lambda\lambda'}(k)=\delta_{\lambda\lambda'}k^2-k_\lambda k_{\lambda'} \ee
and
\be\label{Sigk} \Sigma_{\lambda}(k)=2 G(p) (p-k)_{\lambda'}G_{\lambda'\lambda}(p-k) .\ee
For the charged component we get
\be\label{Pip} \Pi_{\lambda\lambda'}(p)p_{\lambda'}=K_{\lambda\lambda'}(p)\Sigma_{\lambda'}(p)
\ee
with \be\label{Kp}  K_{\lambda\lambda'}(p)=\delta_{\lambda\lambda'}p^2+2iB_{\lambda\lambda'}-p_\lambda p_{\lambda'} \ee
and
\be\label{Sigka} \Sigma_{\lambda}(p)=- G(p-k) \ (p-k)_{\lambda'}G_{\lambda'\lambda}(k)
- G(k) \ k_{\lambda'}G_{\lambda\lambda'}(p-k) .\ee

%%%%%%%%%%%%%%%%%%%%%%%%%%%%%%%%%%%%%%%%%%%%%%%%%%%%%%%%%%%%%%%%%%%%%%%%%%%%%%
%%%%%%%%%%%%%%%%%%%%%%%%%%%%%%%%%%%%%%%%%%%%%%%%%%%%%%%%%%%%%%%%%%%%%%%%%%%%%%
%%%%%%%%%%%%%%%%%%%%%%%%%%%%%%%%%%%%%%%%%%%%%%%%%%%%%%%%%%%%%%%%%%%%%%%%%%%%%
%%%%%%%%%%%%%%%%%%%%%%%%%%%%%%%%%%%%%%%%%%%%%%%%%%%%%%%%%%%%%%%%%%%%%%%%%%%%%%
\section{Dependence of the polarization tensor on the gauge parameter $\xi$}\label{sect4}

The general structure of the r.h.s of the Slavnov-Taylor identity was established by formula \Ref{Wicoord}.
Strictly speaking, it simply states that the 'weak' transversality \Ref{weak} holds. But by knowing the explicit form of $\Sigma^{ef}_\nu(x,z) $, \Ref{avSi}, we are able to calculate the nontransversality explicitly and to judge whether it is zero as for instance without magnetic field or not as in our case. This is the reason for the effort to transform \Ref{Wicoord} into the charged basis and further into momentum representation which resulted in the formulas $\Sigma_{\lambda}(k)$, \Ref{Sigk}, and $\Sigma_{\lambda}(p)$, \Ref{Sigka}. These are simple one loop graphs which can be calculated with the same methods as in \cite{Bordag:2005br} or as in the next section.

By the formulas \Ref{Sigk} and \Ref{Sigka} we also confirmed the
corresponding expressions (118) and (170) in
\cite{Bordag:2005br}. There the same quantities,
$\Sigma_{\lambda}(k)$ and $\Sigma_{\lambda}(p)$, have been
calculated for $\xi=1$ starting from the corresponding one-loop
graphs. In fact, using obvious symmetry properties it is easy to
see that \Ref{Sigk} is the same as (118) in \cite{Bordag:2005br}
and \Ref{Sigka} is coincides with (170) in \cite{Bordag:2005br}.

%%%%%%%%%%%%%%%%%%%%%%%%%%%%%%%%%%%%%%%%%%%%%%%%%%%%%%%%%%%%%%%%%%%%%%%%%%%%%
%%%%%%%%%%%%%%%%%%%%%%%%%%%%%%%%%%%%%%%%%%%%%%%%%%%%%%%%%%%%%%%%%%%%%%%%%%%%%%
\subsection{Neutral part of the polarization tensor}

The neutral polarization tensor has the following representation
in momentum space (cf. Eq.(48) in \cite{Bordag:2005br})
\bea\label{CPT}
\Pi_{\la\la'}(k)&=&\G_{\mu\nu\la}G_{\mu\mu'}(p)\G_{\mu'\nu'\la'}G_{\nu'\nu}(p-k)-
\nn \\ &&p_\la G(p)(p-k)_{\la'}G(p-k)-(p-k)_\la G(p)p_{\la'}G(p-k),
\eea
where the integration over momentum $p$ is assumed. The second
line is the contribution from the ghost loop and
the vertex factor reads
\be \label{vertex}
\G_{\mu\nu\la}=g_{\mu\nu}(k-2p)_{\la}+g_{\la\mu}(p+k)_\nu+g_{\la\nu}(p-2k)_{\mu}.
\ee
The gauge parameter $\xi$ appears in the gluon propagator
\be \label{propxi} G_{\mu\mu'}(p)=\left(g_{\mu\mu'} p^2 -
\left(1-\frac{1}{\xi}\right)p_{\mu}p_{\mu'}+2iF_{\mu\mu'}\right)^{-1},
\ee
which can be represented in the form
\bea \label{propxi1}
G_{\mu\mu'}(p)&=&\left(\frac{1}{p^2+2iF}\right)_{\mu\mu'}-p_{\mu}\frac{1-\xi}{p^4}p_{\mu'} \nn \\
&=&\int_0^\infty ds \
e^{-s(p^2+2iF)}\left(E_{\mu\mu'}^{-s}-s(1-\xi)p_{\mu}p_{\mu'}\right).
\eea Here we used the specific form  of the matrix
$F_{\mu\mu'}=\ep_{\mu\mu' 03}$ of the field strength of the
homogeneous magnetic background field and we introduced the
projectors $\delta^{\parallel}_{\mu\mu'}$ and
$\delta^{\perp}_{\mu\mu'}$ onto the subspaces parallel and
perpendicular to the magnetic field. The exponential in the
representation of the propagator as parametric integral is
\be
E_{\mu\mu'}^{s}\equiv(e^{2isF})_{\mu\mu'}=\delta^{\parallel}_{\mu\mu'}-iF_{\mu\mu'}\sinh
(2s)+\delta^{\perp}_{\mu\mu'}\cosh (2s).\ee

For $\xi=1$ we did the calculations in \cite{Bordag:2005br}. Here we
calculate the $\xi$-dependent contributions which appear proportional  to $(1-\xi)$ and to
$(1-\xi)^2$.

In the following we will make use of the simple commutation properties
\be \label{cp} p_{\mu}\left(\frac{1}{p^2+2iF}\right)_{\mu\mu'}=\frac{1}{p^2} \ p_{\mu'} \ , \qquad
\left(\frac{1}{p^2+2iF}\right)_{\mu\mu'}p_{\mu'}=p_{\mu} \ \frac{1}{p^2}\ee
and of the obvious relations
\be \label{Gak}
\begin{array}{rclrcl}
\Gamma_{\mu\nu\la}p_{\mu}&=& K_{\nu\la}(p-k)-K_{\nu\la}(k),&
p_{\mu}\Gamma_{\mu\nu\la}&=& K_{\la\nu}(p-k)-K_{\la\nu}(k), \\
(p-k)_{\nu}\Gamma_{\mu\nu\la}&=& K_{\la\mu}(p)-K_{\la\mu}(k),&
\Gamma_{\mu\nu\la}(p-k)_{\nu}&=& K_{\mu\la}(p)-K_{\mu\la}(k),
\end{array}
\ee
holding for the vertex factors. Here the $K$'s are given by Eqs. \Ref{Kk} and \Ref{Kp}.

In order to make the following calculations easier to follow we
use the more symbolic notation for the propagators as inverse
powers. Again, the momentum integration is not shown. In this way
inserting \Ref{propxi} into \Ref{CPT} we write the polarization
tensor in the form
\bea
\Pi_{\la\la'}(k)&=&\Pi_{\la\la'}(k)_{\mid{\xi=1}}-(1-\xi)\G_{\mu\nu\la}p_\mu{1
\o p^4}p_{\mu'}\G_{\mu'\nu'\la'}\left({1 \o
(p-k)^2+2iF}\right)_{\la\la'} \nn \\ && -
(1-\xi)\G_{\mu\nu\la}\left({1 \o
p^2+2iF}\right)_{\la\la'}\G_{\mu'\nu'\la'}(p-k)_{\nu'}{1 \o
(p-k)^4 }(p-k)_\nu  \nn \\ && + (1-\xi)^2\G_{\mu\nu\la}p_\mu{1 \o
p^4}p_{\mu'}\G_{\mu'\nu'\la'}(p-k)_{\nu'}{1 \o (p-k)^4}(p-k)_\nu .
\eea
Further we note that the whole expression is under a trace
with respect to the operators $p_\mu$ and that cyclic permutations are
possible. Making use of the mentioned properties some
simplifications can be made, and  the expression for the
polarization tensor can be represented as
\bea \label{struck} \Pi_{\la\la'}&=& \Pi_{\la\la'}(\xi=1)
-(1-\xi) \left(
K_{\la\nu}(k)\Sigma^1_{\nu\la'}+\Sigma^1_{\la\nu'}K_{\nu'\la'}(k)
\right. \nn \\ &&  \left.
+K_{\la\nu}(k)\Sigma^2_{\nu\nu'}K_{\nu'\la'}\right) +(1-\xi)^2 \
K_{\la\nu}(k)\Sigma^3_{\nu\nu'}K_{\la'\nu'}(k), \eea
with the new notations
\bea \Sigma^1_{\nu\la}&=&(p-k)_\nu{1 \o p^4}(p-k)_{\la}{1 \o
(p-k)^2}+p_\nu {1 \o p^2}p_{\la}{1 \o (p-k)^4 } \ ,\nn \\
\Sigma^2_{\nu\nu'}&=&{1 \o p^4}\left({1 \o
(p-k)^2+2iF}\right)_{\nu\nu'}+\left({1 \o
p^2+2iF}\right)_{\nu\nu'}{1 \o (p-k)^4}  \ , \nn \\
\Sigma^3_{\nu\nu'}&=&p_\nu{1 \o p^4 }p_{\nu'}{1 \o (p-k)^4} \
.\eea
Note that the loop integration over $p$ is only inside these
quantities. As in \cite{Bordag:2005br} we use Schwingers' operator method
where the loop integration is rewritten as a specific averaging
and the result is represented as a double parametric integral,
see Eq. (60) in \cite{Bordag:2005br}. Doing the same for the $\Sigma$'s we
write
\be \label{SigM}
\Sigma^{i}_{\la\la'}=\int_0^\infty\int_0^\infty \ ds \ dt \
\langle M^{\Sigma^i}_{\la\la'}(p,k) \Theta \rangle  .  \ee
The corresponding functions $M$ are
\bea M^{\Sigma_1}_{\nu\la}(p,k)&=&s \ (p-k)_\nu(p(s)-k)_{\la}+t\ p_\nu
p(s)_{\la} \ , \nn \\
M^{\Sigma_2}_{\nu\nu'}(p,k)&=&sE^{-t}_{\nu\nu'}+tE^{-s}_{\nu\nu'} , \nn
\\ M^{\Sigma_3}_{\nu\nu'}(p,k)&=&st p_\nu p(s)_{\nu'}
\eea
with $ p(s)_\mu\equiv e^{-sp^2}p_\mu e^{sp^2}={E^s}_{\mu\nu} p_\nu $.
The averaging is given by the formulas (55), (56) and (57) in \cite{Bordag:2005br} and by means of $\langle M_{\la\la'}(p,k) \Theta \rangle = M_{\la\la'}(s,t) \langle
\Theta \rangle$ we arrive at
\bea M^{\Sigma_1}_{\nu\la}(s,t)&=&s \ P_\nu P_\la^T+t \ Q_\nu Q_\la^T
+(s+t)\left(  a \ \delta_{\nu\la}^{||}+b \  i  F_{\nu\la}+c\
\delta_{\nu\la}^{\perp} \right)  ,
\nn \\
M^{\Sigma_2}_{\nu\nu'}(s,t)&=&
(s+t)\delta_{\nu\la}^{||}-(s \st + t \st) i  F_{\nu\la}+(s\ct+t\cs)\delta_{\nu\la}^{\perp},
\nn \\ &\equiv &
\tilde{a} \ \delta^{\parallel}_{\nu\nu'}+ \tilde{b} \ iF_{\nu\nu'}+\tilde{c} \ \delta^{\perp}_{\nu\nu'}  ,
\nn \\
 M^{\Sigma_3}_{\nu\nu'}(s,t)&=&st \ Q_\nu Q^T_{\nu'}
+a\ \delta_{\la\la'}^{||}+b\ iF_{\la\la'}+c\
\delta_{\la\la'}^{\perp},
 \eea with the notations \bea P_\la&=&r\
l_\la+ \al\  i d_\la+\b \ h_\la , \nn \eea\vspace{-12pt}
\be\begin{array}{rclrclrcl} r&=&{s \o s+t}, &
\al&=&-{\sinh(s)\sinh(t) \o \sinh (s+t)}, & \b&=&{\cosh(s)
\sinh(t) \o \sinh (s+t)} ,
\end{array}\ee
\bea
Q_\la&=&P_{\la}(s\leftrightarrow t) \nn\ \\
&\equiv&  s' \ l_\la+ \gamma \  i d_\la+\delta \ h_\la , \nn
\eea
and
\be\label{abc}a={1\o s+t}, \quad b=-{\sinh(s-t) \o \sinh(s+t)}, \quad c={\cosh(s-t) \o \sinh(s+t)}.
\ee
The vectors $l_\la$, $h_\la$ and $d_\la$ are
given by Eqs. (38) in \cite{Bordag:2005br}.

These expressions can be rewritten as contributions to the formfactors which are introduced by means of the decomposition
\be \Pi_{\la\la'}(p)=\sum_{i=1}^6
\Pi^{(i)}(l^2,h^2+2iF)_{\la\la''}T_{\la\la''}^{(i)}\ee with the
operators $T_{\la\la''}^{(i)}$ given by Eq. (39) in
\cite{Bordag:2005br}. In the corresponding expression in form of
double parametric integrals, \be\label{formfM}
\Pi^{(i)}(k)=\int_0^\infty\int_0^\infty \ ds \ dt \ M^{(i)}(s,t)
\ \langle\Theta\rangle, \ee the functions $M^{(i)}(s,t)$ of the
$\xi$-dependent part reads
\bea\label{Mineut}
M^{(1)}(s,t)&=&-(1-\xi)[t(-2rh^2x)+4(s+t)+s(-2s'h^2x')+\tilde a
(s+t)+\tilde c(-h^2)]
     \nn \\&& \qquad  +(1-\xi)^2 st[a(l^2+2h^2)+c(-h^2)], \nn \\
M^{(2)}(s,t)&=&-(1-\xi)[t(2l^2\b x+2\al^2(l^2+h^2))+s(2l^2\delta
x'+2\gamma^2(l^2+h^2))+\tilde a(-l^2)\nn \\&& \qquad+\tilde c
(2l+h^2)]
       +(1-\xi)^2[al^2+c(2l^2+h^2)] ,\nn \\
M^{(3)}(s,t)&=&-(1-\xi)[t(-\b h^2+rl^2)x+2(s+t)+s(s'l^2-\delta
h^2)x'+\tilde a h^2+\tilde c l^2]
     \nn \\&&  \qquad -(1-\xi)^2[ah^2+cl^2] , \nn \\
M^{(4)}(s,t)&=&-(1-\xi)[t(-\al)(h^2x+r(l^2+h^2))+s(-\al)(h^2x'+s'(l^2+h^2))+\tilde
b  (l^2+h^2)]
     \nn \\&&  \qquad +(1-\xi)^2[b(l^2+h^2)],\nn \\
M^{(5)}(s,t)&=&-(1-\xi)[t x(rl^2+\b h^2)+s x'(s'l^2+\delta h^2)] ,\nn \\
M^{(6)}(s,t)&=&-(1-\xi)[t\al (l^2+h^2)(rl^2+\b h^2)+s\gamma
(l^2+h^2)(s'l^2+\delta h^2)] . \eea with the notations $x=r-\b$
and $x'=s'-\delta$. These functions must be added to the
corresponding ones in formula (103) in \cite{Bordag:2005br} to get the
complete $\xi$-dependent polarization tensor.

\subsection{Charged part of the polarization tensor}

In the same manner can be done the calculations for
$\xi$-depending part of the charged polarization tensor. We start
from the following representation of charged polarization tensor
in momentum space (cf. Eq.(139) in \cite{Bordag:2005br})
\bea
\Pi_{\la\la'}(p)=\G_{\la\nu\rho}G_{\nu\nu'}(p-k)\G_{\la'\nu'\rho'}G_{\rho\rho'}(k)+\nn \\
(p-k)_\la G(p-k)k_\la'G(k)+k_\la G(p-k)(p-k)_{\la'}G(k), \eea
where the integration over momentum $k$ is assumed. The second
line is the contribution from the ghost loop. The vertex factor
is the same as (\ref{vertex}) but with renamed indices
\be \Gamma_{\la\nu\rho}=(k-2p)_\rho
g_{\la\nu}+g_{\rho\nu}(p-2k)_\la+g_{\rho\la}(p+k)_\nu \ee
Again using (\ref{propxi}), commutation propreties (\ref{cp}) and
relations (\ref{Gak}) we write polarization tensor in the form
\bea \Pi_{\la\la'}(p)=\Pi_{\la\la'}(p)_{\mid{\xi=1}}-(1-\xi)\left[
K_{\la\rho} (p) \Sigma^1_{\rho\rho'}K_{\rho\la'}+ \right. \nn \\
\left. K_{\la\rho}(p)\Sigma^2_{\rho\la'}
+\Sigma^3_{\la\rho}K_{\rho\la'}(p)\right]+(1-\xi)^2\left[
K_{\la\rho}(p)\Sigma^4_{\rho\rho'} K_{\rho'\la'}(p)\right],
\l{pol_tens} \eea
with the notations
\bea \Sigma^1_{\rho\rho'}&=&{1 \o (p-k)^4}{\delta_{\rho\rho'} \o
k^2}+{1 \o k^4}{\delta_{\rho\rho'}\o (p-k)^2}, \nn \\
\Sigma^2_{\rho\la'}&=&{1 \o (p-k)^4}{1 \o k^2}k_\rho k_{\la'}-{1
\o k^4}{1 \o (p-k)^2}k_\rho (p-k)_{\la'}, \nn \\
\Sigma^3_{\la\rho}&=&{1 \o (p-k)^4}{1 \o k^2}k_\la k_\rho-k_\rho
(p-k)_{\la}{1 \o k^4}{1 \o (p-k)^2}, \nn \\
\Sigma^4_{\rho\rho'}&=&{1 \o (p-k)^4}{1 \o k^4} k_\rho k_{\rho'}.
\eea
The momentum integration is not shown and whole expression is
under a trace with respect to operators $k$.
It is straightforward from equation (\ref{pol_tens}) that the
polarization tensor does not depend on gauge parameter $\xi$.
The corresponding functions $M$ are
\bea M_{\rho\rho'}^{\Sigma_1}&=&(s+t)\delta_{\rho\rho'}, \nn \\
M_{\la'\rho}^{\Sigma_2}&=&sk_\rho k_{\la'} -t k_\rho (p-k)_{\la'}, \nn \\
M_{\la\rho}^{\Sigma_3}&=&s k_\la k_\rho -tk_\rho (p-k)_{\la}\nn, \\
M_{\rho\rho'}^{\Sigma_4}&=&stk_\rho k_{\rho'}.\eea
After the averaging and by means of $\langle M_{\la\la'}(p,k)
\Theta \rangle = M_{\la\la'}(s,t) \langle \Theta \rangle$ we get
for functions $M$
\bea M^{\Sigma_1}_{\rho\rho'}&=&(s+t)(\delta^\parallel_{\rho\rho'}+\delta^\perp_{\rho\rho'}), \nn \\
M^{\Sigma_2}_{\rho\la'}&=&sP_\rho P^T_{\la'}-tP_\rho
Q_{\la'}+(a\delta^{||}+biF+c\delta^{\perp})(s+t), \nn \\
M^{\Sigma_3}_{\la\rho}&=&s P_{\la}P^T_\rho+t Q^T_\la
P^T_\rho+(a\delta^{||}+biF+c\delta^{\perp})(s+t), \nn \\
M^{\Sigma_4}_{\nu\nu'}&=&st[P_\nu
P^T_{\nu'}+a\delta_{\la\la'}^{||}+biF_{\la\la'}+c\delta_{\la\la'}^{\perp}]
\cqq \eea
with the notations
\bea P_\la &=& rl_\la+ \al i d_\la+\b h_\la, \nn \\
Q_\la &=& s' l_\la+\gamma i d_\la+\delta h_\la, \nn \\ a&=&{1 \o
2} {1 \o s+t} \cqq \nn \\ b&=&{1 \o 2} \left({\cosh (2s) (\cosh
(2s)
-1) \o 2N}+{\sinh (2s) (\sinh (2s) +2t) \o 2N}\right) \cqq \nn \\
c&=&{1 \o 2} \left( {2 t \cosh (2s) +\sinh (2s) \o 2N} \right),
\eea
and
\be\begin{array}{rclrclrcl} \\r&=&{s \o s+t}, \qq \al&=&{t (\cosh
(2s)-1) \o 2N}, \qq \b&=&{\cosh (2s)-1+t\sinh (2s) \o 2N} \nn \\
s'&=&{t \o s+t}, \qq \gamma &=&{t (\cosh (2s)-1) \o 2N}, \qq
\delta &=& 1-{\cosh (2s)-1+t\sinh (2s) \o 2N}, \end{array}\ee
with $N={1 \o 4}[(\sinh (2s)+2t)^2-\cosh (2s)-1)^2]$.

Finally we present the expression as contributions to the
formfactors. The functions $M^{(i)}(s,t)$ of the $\xi$-dependent
part read
\bea \label{Michar}
M^{(1)}&=&-(1-\xi) [t(2s'(xH^2 +\al))+2rs(xH^2+\al) + 2 a(s
+t)+(s + t) (l^2 + H^2)] + \nn \\&& \qq (1 - \xi)^2 [-st(x H^2 +
\al)^2 + a (l^2 + 2 H^2) + b - c H^2)], \nn \\
M^{(2)}&=&-(1- \xi)
[-2 t (\delta (x l^2 - \al) - \al \gamma(l^2 + H^2)) + \nn \\
&& 2 s (\beta (x l^2 - \al) - \al^2 (l^2 + H^2))+(s+t)(l^2+H^2)+2
c(s + t)]  \nn \\ && + (1 - \xi)^2 [s t (-(x l^2 -
\al)^2+\al^2(l^2 + H^2)^2) + \nn \\ && s t (-a l^2 + b + c(2l^2 +
H^2))], \nn \\
M^{(3)}&=&-(1 - \xi) [-t(s'(x l^2 - \al - \delta (x H^2 + \al)) +
s(r(x l^2 - \al) - \al (x H^2 + \al)) + \nn
\\ && (s + t) (a + c)+(s + t)(l^2 +
H^2)] + \nn \\ && (1 - \xi)^2 [s t (x l^2 - \al) (x H^2 + \al) + s t (a H^2 + b + c l^2)],  \nn \\
M^{(4)}&=& -(1 - \xi) [-t (\gamma (x H^2 + \al) - \al s' (l^2 +
H^2)) +s (\al (x H^2 + \al) \nn \\&& - \al r(l^2 + H^2) +(s + t)
b]  + (s + t)(l^2 + H^2) \nn \\&& (1 - \xi)^2 [s t (-\al)(l^2 +
H^2)(x H^2 +\al)+s t b (l^2 + H^2)], \nn \\
M^{(5)}&=&-(1-\xi) [-t x(s l^2 + \delta H^2 + \gamma)+s  x(r l^2
+ \beta H^2 +\al) +(s + t) (a
- c) ],\nn \\
M^{(6)}&=&-(1-\xi) [-t\al (s l^2 + \delta H^2 + \gamma) + s \al (r
l^2 + \beta H^2 + \al) + (s + t) b], \eea
with $x=r-\b$.

%%%%%%%%%%%%%%%%%%%%%%%%%%%%%%%%%%%%%%%%%%%%%%%%%%%%%%%%%%%%%%%%%%%%%%%%%%%%%%
%%%%%%%%%%%%%%%%%%%%%%%%%%%%%%%%%%%%%%%%%%%%%%%%%%%%%%%%%%%%%%%%%%%%%%%%%%%%%%
%%%%%%%%%%%%%%%%%%%%%%%%%%%%%%%%%%%%%%%%%%%%%%%%%%%%%%%%%%%%%%%%%%%%%%%%%%%%%%
\section{Conclusions}
In the foregoing sections we showed that the gluon polarization tensor in a chromomagnetic background field is non transversal and in addition we calculated its dependence on the gauge fixing parameter $\xi$. It should be remarked that the non-transversality is by no means in conflict with the gauge invariance of the model. It simply signals a more complicated structure of the Greens functions.  As a by-product the derivation of the non-transversal parts in section 2 from the Slavnov-Taylor identities justified the calculation in \cite{Bordag:2005br} where the same result was obtained  from the one-loop graph in a calculation  where the tadpole graphs had been omitted. In the calculation in section 2  nothing was omitted and in this way it was confirmed that the tadpole graphs add up to zero.

The technique for the calculation of the one-loop graphs used in \cite{Bordag:2005br} which is based on Schwinger's operator method was used to calculate the $\xi$-dependent part of the polarization tensor. The  dependence on $\xi$ is non-trivial and the final formulas look even more complicated as that for $\i=1$, see \Ref{Mineut} resp. \Ref{Michar} in comparison with (103 resp. (164) and (177) in \cite{Bordag:2005br}.

Because, of course, a physical final result should not depend on the gauge fixing parameter $\xi$ it should be sufficient to perform all calculations in the easier case of $\xi=1$. However, having in mind partial resummations in order to improve the infrared properties or to investigate such effects as spontaneous generation of condensates the question about the $\xi$-dependence becomes less simple. The point is that in such procedures the gauge invariance may be broken in general, although keeping it unbroken in the given approximation. In such cases the knowledge of the $\xi$-dependence can be helpful.

Eventually we demonstrate on the pure one loop level that a physical result, namely the expectation value of the polarization tensor in physical states does not depend on $\xi$. The states are written down explicitly in \cite{Bordag:2005br}. The neutral states are denoted by $\mid k_{||}, k_\perp,s\rangle_\mu$ and the charged ones by $\mid p_{||}, n,s\rangle_\mu$, where  the momenta are divided into the components parallel and perpendicular to the magnetic field, $p_\perp$  appears in form of the Landau level $n$ and  $s$ ($s=1,\dots,4$) denotes the polarizations which for vanishing magnetic field coincide with the standard photon polarizations, $s=1,2$ for the transversal ones, $s=3$ for the longitudinal one and $s=4$ for the time-like one (after going back into   Minkowski space). These states have the properties
\bea
K_{\mu\nu}(k)  \ \mid k_{||}, k_\perp,s>_\nu&=&(k_4^2+k^2) \ \mid k_{||}, k_\perp,s>_\mu  \nn \\
K_{\mu\nu}(p) \ \mid p_{||},n,s\rangle_\nu &=& \left(p_{||}^2+B(2n+1) \right)\mid p_{||},n,s\rangle_\mu
\eea
for $s=1,2$ where the $K_{\mu\nu}$ are the operators \Ref{Kk} and \Ref{Kp}. Now we take into account the structure of the $\xi$-dependent part as given by Eqs. \Ref{struck} for the neutral and by \Ref{pol_tens} for the charged component.

If we now calculate the expectation value
\be\label{expvalue}
{}_\mu\langle k_{||}, k_\perp,s \mid \Pi_{\mu\nu}(k) \mid k_{||}, k_\perp,t>_\nu
\ee
or the corresponding quantity for the charged component than for $s,t=1,2$ (\Ref{expvalue} does not need to be diagonal in $s$ and $t$) the $\xi$-dependent part is proportional to $ (k_4^2+k^2) )$ or to $\left(p_{||}^2+B(2n+1) \right)$. Now it remains to remark that after returning to Minkowski-space these factors vanish on-shell. In this way we  showed  that the physical on-shell matrix elements  of the polarization tensor  do not depend on the gauge fixing parameter $\xi$.

A remark is in order. The one loop matrix elements contain on-shell an infrared divergence and in QCD they do not have direct physical significance. So this result is of rather methodological interest. On the other hand side the structure of the $\xi$-dependent part  found in \Ref{struck} and \Ref{pol_tens} is not trivial as can be seen from the $\xi=1$-pert which is weakly transversal too but which gives a non vanishing contribution to the mentioned matrix elements.

\section*{Acknowledgement}

The authors JG and VS thank he Institute for Theoretical Physics of Leipzig university  for kind hospitality, JG thanks the DAAD for support through the Euler-program and VS thanks the NTZ for support.

\bibliographystyle{EJPC}\bibliography{../Text0507141/PolTensorSU2,../../../Literatur/Bordag}
 \end{document}